# Terabit optical OFDM superchannel transmission via coherent carriers of a hybrid chip-scale soliton frequency comb

Yong Geng,[1] Xiatao Huang,[1] Wenwen Cui,[1] Yun Ling,[1] Bo Xu,[1] Jin Zhang,[1] Xingwen Yi,[1] Baojian Wu,[1] Shu-Wei Huang,[2] Kun Qiu,[1] Chee Wei Wong,[3,*] and Heng Zhou[1,*]

[1]*Key Lab of Optical Fiber Sensing and Communication Networks, School of Information and Communication Engineering, University of Electronic Science and Technology of China, Chengdu 611731, China*
[2]*Department of Electrical, Computer, and Energy Engineering, University of Colorado, Boulder, Colorado 80309, USA*
[3]*Fang Lu Mesoscopic Optics and Quantum Electronics Laboratory, University of California, Los Angeles, CA 90095, USA*
*\*Corresponding author: cheewei.wong@ucla.edu; zhouheng@uestc.edu.cn*

**We demonstrate seamless channel multiplexing and high bitrate superchannel transmission of coherent optical orthogonal-frequency-division-multiplexing (CO-OFDM) data signals utilizing a dissipative Kerr soliton (DKS) frequency comb generated in an on-chip microcavity. Aided by comb line multiplication through Nyquist pulse modulation, the high stability and mutual coherence among mode-locked Kerr comb lines are exploited for the first time to eliminate the guard intervals between communication channels and achieve full spectral density bandwidth utilization. Spectral efficiency as high as 2.625 bit/Hz/s is obtained for 180 CO-OFDM bands encoded with 12.75 Gbaud 8-QAM data, adding up to total bitrate of 6.885 Tb/s within 2.295 THz frequency comb bandwidth. Our study confirms that high coherence is the key superiority of Kerr soliton frequency combs over independent laser diodes, as a multi-spectral coherent laser source for high-bandwidth high-spectral-density transmission networks.**

http://dx.doi.org/unpublished

Kerr frequency combs produced in high-quality factor nonlinear microcavities present a new and promising platform for versatile multi-tone multi-color light carriers for frequency synthesis [1-5], precision metrology [6-7], laser spectroscopy [8-9], time keeping [10-11], microwave and terahertz signal generation [12-14], and optical communications [15-16]. These functionalities of Kerr frequency comb furthermore can be achieved in a compact chip-scale footprint, with simultaneous broadband spectrum, high signal-to-noise ratio, high precision and stability, and low phase noise at the thermodynamical limits. In particular, such integratable frequency combs constitute an intrinsic multi-channel laser source for terabit optical transmission networks [17-20]. Various types of Kerr combs including phase-locked microcomb associated with stable Turing pattern [17-19], and mode-locked combs associated with bright dissipative Kerr soliton (DKS) [20] as well as dark solitons, have recently been leveraged to implement ultrahigh-capacity optical transmissions, with record bitrate in excess of 50 Tb/s recently achieved by two interleaved DKS combs covering the entire optical communication C- and L-bands [19].

However, one of the most important and advantageous traits of Kerr frequency microcomb as multichannel electromagnetic carriers – the high mutual coherence (line-to-line spacing stability) among all comb lines -- has yet to be exploited in data transmission. This can potentially allow the long-sought zero inter-channel guard interval operation [20], along with deterministic nonlinear compensation [22]. The challenges for this testbed realization are multifold. First, to generate the Kerr comb with the mode spacing compatible with the bandwidth of the state-of-art electronic data transceiver modules (e.g. 10 to 40 GHz), it requires relatively large diameter microcavities, which in turn leads to increased fabrication complexity and higher comb parametric oscillation threshold [23]. Second and more importantly, because of the current low power conversion efficiency from the pump laser to Kerr comb lines [24], small comb spacings can result in reduced comb power per line, thus making it tougher for subsequent comb amplifications and coherent modulation, wherein the laser carriers should have high input power and optical-signal-to-noise-ratio (OSRN) as high as possible. Previously reported efforts to reduce the Kerr comb spacing (so as to achieve higher transmission spectral efficiency) include interleaving two identical DKS combs [20], and inserting extra comb lines via electro-optical (EO) modulation [25]. Nevertheless, these prior demonstrations have not completely close up the intervals between data channels, and whether the mutual coherence of Kerr comb can support seamless channel stitching and superchannel transmission still remains an open question.

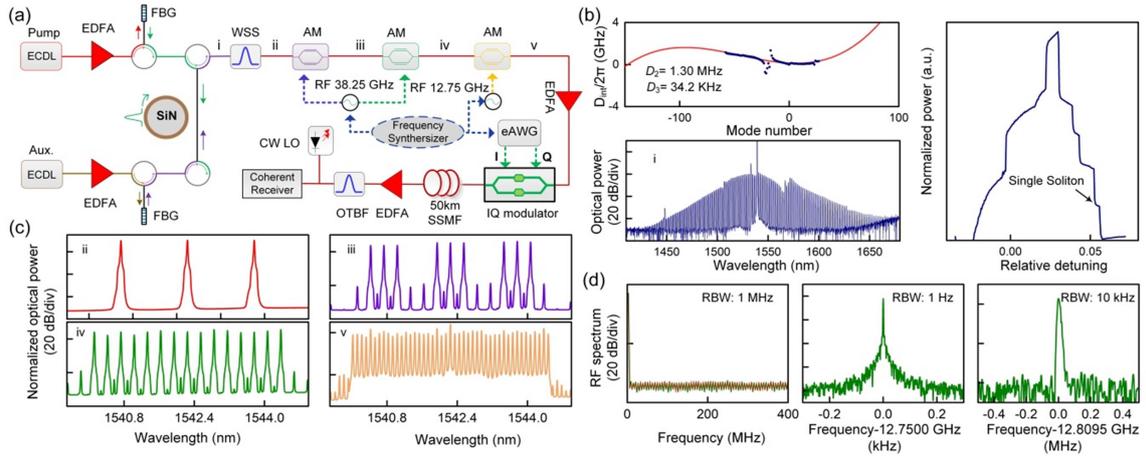

Fig. 1. The generation and characterization of Kerr/EO frequency comb. (a). Experiment setup for DKS comb generation, Nyquist pulse modulation, and CO-OFDM transmission. (b). Right: optical spectrum of a single DKS microcomb (lower); and the corresponding cavity dispersion (upper), blue-dot is from measurement and red-line is from FDTD simulation. Left: experimentally measured soliton steps. (c). Optical spectrum of ii) 3 initial Kerr comb lines; iii) 9 comb lines after the first MZM; iv) 15 comb lines after the second MZM; v) 45 comb lines after the third MZM. (d). Left panel: the corresponding RF spectrum at the detector noise floor. Middle panel: the beat note measured for the EO modulated comb lines. Right panel: the beat note of the edge comb lines from neighboring EO comb families (the 15th and 16th comb line). ECDL: external-cavity diode laser; EDFA: erbium doped fiber amplifier; FBG: fiber Bragg grating; WSS: wavelength selective switches; AM: amplitude modulator; OTBF: optical tunable bandpass filter.

Here we demonstrate for the first time dense spectra bundling of coherent optical orthogonal-frequency-multiplexing (CO-OFDM) signals using an actively-stabilized on-chip DKS microcomb. Through Nyquist pulse modulation [27] and delicately deployed optical amplification, we successfully reduce the comb spacing from 191.31 GHz to 12.75 GHz while still maintaining high OSNR of the individual comb lines and with low inter-line power fluctuation. The reduplicated comb lines are used to carry 12.75 Gbaud 8-QAM CO-OFDM data signals, achieving seamless spectra aggregation and ultrahigh bitrate superchannel transmission empowered by the high stability of mode-locked Kerr frequency comb.

Figure 1(a) shows the experimental setup for the chip-scale frequency comb generation. A silicon nitride microring resonator with cross-section of 800×2000 $nm^2$ and loaded $Q$-factor of $\approx$ 500,000 is used as the nonlinear microcavity. Measured cavity dispersion is shown in Figure 1(b), revealing an anomalous GVD around the pump wavelength. An auxiliary-laser-heating method is adapted to access and actively stabilize single DKS Kerr comb state [26]. The basic idea of our method is to launch an auxiliary laser into one cavity mode (~1532.8 nm) and preset its wavelength within the blue-detuning regime. Then, a c.w. pump laser is counter-propagating launched into the microcavity and its frequency tuned into another cavity mode (~1538.8 nm). As the pump laser enters the cavity mode from the blue-detuning range, the microcavity is heated and all the resonances are thermally red-shifted. However, the red-shifted resonances displace the auxiliary laser from its own cavity mode, in tune cooling the cavity. By properly setting the power and detuning of pump and auxiliary lasers, the heat flow incurred by them tend to balanced out each other. This allows the pump laser to scan across the entire cavity resonance linewidth without thermal dragging. By implementing the auxiliary-laser-heating scheme, we can stably generate single DKS microcomb with high repeatability. The generated DKS comb spectrum is shown in Figure 1(b), with a mode spacing of $\approx$ 191.31 GHz for the fundamental $TM_{00}$ mode. The RF spectrum of the generated Kerr comb is also shown in Figure 1(d), exhibiting low amplitude noise and confirming the mode-locked DKS state. A 17 nm C-band portion (12 comb lines from 1540.704 to 1557.552 nm) of the DKS comb is filtered as the initial comb lines for subsequent processing.

Importantly, thanks to the high repetition rate of the DKS pulses, each of the selected comb line has a power of about -20 dBm, which is high enough to maintain their OSNR to more than 35 dB after being boosted to +5 dBm using a low-noise EDFA. We note that this 17 nm comb spectrum bandwidth is limited solely by the power budget and noise figure of our optical amplification instrumentation. In fact, the generated Kerr comb achieves decent power levels spanning the entire C- and L-band, all of which could be utilized to carry high bitrate data signals by virtue of more amplifiers [20].

The 191.31 GHz mode spacing of the initial comb lines is far beyond the bandwidth of our CO-OFDM transceiver. So we subsequently send the initial comb lines to a Nyquist pulse generator consisting of three Mach-Zehnder modulators (MZM) [27]. Two sinusoidal RF signals at 38.25 GHz and 12.75 GHz from two synchronized microwave synthesizers are used to drive these MZMs, with the configuration plotted in Figure 1(a). By carefully adjusting the RF amplitude and bias voltage applied to each MZM, the Nyquist pulse trains are modulated onto the initial comb lines, producing multiple evenly-distributed EO modulation sidebands. To warrant favorable Nyquist pulse modulation, we modulate three Kerr comb lines at a time and generate 15 sidebands around each of them, giving rise to 45 hybrid Kerr-EO comb lines with overall flat spectrum envelope, as exemplified in Figure 1(c). Note that the 191.31 GHz Kerr comb spacing offsets slightly from the 15×12.75 GHz = 191.25 GHz overall EO modulation span, resulting in a 12.8095 GHz spacing between the two most edge sidebands from two adjacent EO comb families. The reason for this discrepancy and its impact on CO-OFDM performance will be discussed later in this study. Importantly, it is seen that the hybrid Kerr-EO comb lines all maintain good OSNR better than 20 dB.

Figure 1(d) characterizes the mutual coherence of the generated hybrid Kerr-EO comb. First, the EO comb lines obtained via Nyquist pulse modulation of initial Kerr comb lines are locked to the driving RF signals derived from stable microwave synthesizers. As shown in the middle panel of Figure 1(d), the beat note of these modulated sidebands is narrower than the resolution of our electrical spectrum analyzer (ESA, illustrating their inherent high coherence.) Second, the DKS comb coherence is examined by recording the 12.8095 GHz beat note produced by the edge modulation sidebands. As shown in the right panel of Figure 1(d), the beat note also exhibits a narrow linewidth less than 50 kHz, perturbed by the amplitude and frequency jitter of the free-running pump and auxiliary lasers [13, 28]. Therefore, the generated hybrid Kerr-EO comb now has compatible mode spacing with our data transceiver, as well as excellent mutual coherence among all comb lines, enabling us to conduct high bitrate data transmission with superior spectral utilization.

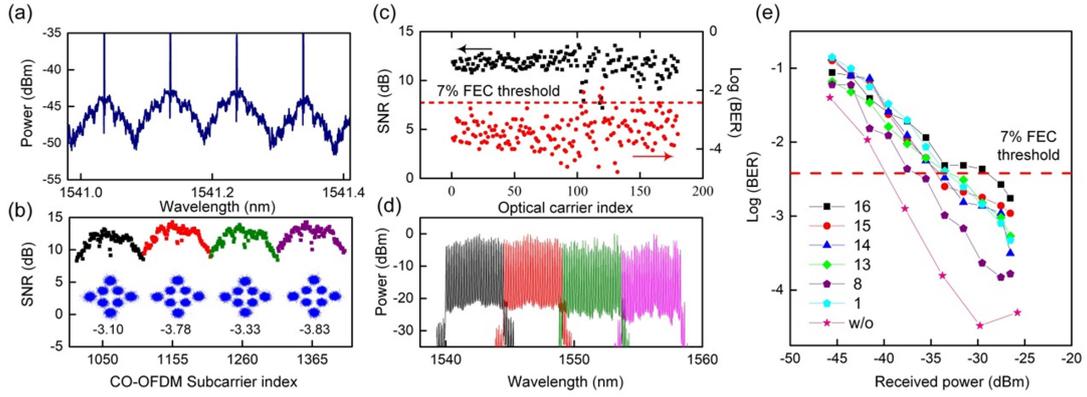

Fig. 2. Terabit seamless CO-OFDM data transmission experiment using hybrid Kerr-EO frequency comb with 180 coherent channels as a prototype. (a). Optical spectra of 4 CO-OFDM bands measured using a stimulated Brillouin scattering-based optical spectrum analyzer. (b). SNR performance for each OFDM subcarriers carried by comb lines 11, 12, 13 and 14 after 50 km SSMF transmission. The inset shows the constellations for each OFDM signal, with log(BER) of -3.10, -3.78, -3.33 and -3.83, as denoted below each constellation. (c). The SNR and BER estimation across all 180 CO-OFDM channels. Red-dotted line denotes the 7% FEC threshold. (d). Optical spectrum of four hybrid Kerr-EO combs containing 180 comb lines, to illustrate the quality of each comb line carrier. (e). BER sensitivity of 6 CO-OFDM bands after 50 km SSMF transmission, in comparison with the transmission performance of one initial Kerr comb line without Nyquist pulse modulation.

Figure 1(a) shows the setup of the CO-OFDM transmission experiment. First, the hybrid Kerr-EO comb (containing 45 lines) is boosted by a preamplifier to about 19 dBm total power (about +2 dBm each line). Then these comb lines are sent into an IQ modulator for CO-OFDM data modulation. The CO-OFDM data symbol is generated within an electrical arbitrary waveform generator (eAWG), via 128-point long fast Fourier transformation (FFT). 102 subcarriers are used to carry the pilot and data information, with the cyclic prefix being 1/16 of the total FFT length [29]. Two measures are taken to ensure that the comb lines serve as orthogonal laser carriers for seamless assembling of the OFDM bands. First, the sampling rate of eAWG is set to 24.0 GS/s, so that the baud rate of the encoded 8-QAM OFDM data is 12.75 Gbaud (102/128×24 GSample/s÷1.5 Sample), consistent with the EO comb line-to-line spacing. Of note, in the EO comb generation we did not equally separate the 191.31 GHz Kerr comb spacing (191.31/15=12.754 GHz), since it will require non-integer sampling rate (24.0075 GS/s) of the eAWG and make the subsequent data decoding algorithm more intractable. Second, the RF signal for Nyquist modulation and the eAWG clock for CO-OFDM data generation are synchronized using a 10 MHz reference clock, so as to avoid orthogonality degradation due to time-base offset. We also note that, unlike field scenarios, in our experiments the 45 comb lines are encoded with identical data for demonstration purposes. This represents the worse-case in terms of power variation and the peak-to average ratio (PAR) of the generated OFDM signals [29], thus will not compromise the validity of our demonstration. Figure 2(a) shows the high-resolution spectrum of 4 CO-OFDM channels measured by a stimulated Brillouin scattering based optical spectrum analyzer. It is clearly seen that all the CO-OFDM bands are seamlessly connected without guard interval, enabling full spectral utilization.

The 45-channel 8-QAM CO-OFDM data signals are transmitted over 50 km standard single-mode fiber (SSMF) and then fed into the CO-OFDM receiver. At the receiver, the incoming signals are combined with a 100 kHz linewidth laser local oscillator (LO) for heterodyne coherent detection. The 8-QAM data carried by each comb lines (selected by the LO frequency) is collected using a 50 GS/s digital-processing oscilloscope (DPO), and decoded using our offline data decoder aided by DSP algorithms including frequency offset compensation, phase noise and channel compensation, and timing synchronization. Figure 2(b) shows the estimated signal-to-noise ratio (SNR) for 4×102 OFDM subcarriers contained in 4 CO-OFDM bands, all exhibiting high signal quality with average SNR better than 11 dB. Slight SNR degradation is observed for those side subcarriers, which is due to the bandwidth limitation of our transmitters and receivers. Nevertheless, each CO-OFDM band achieves low bit-error-rate (BER) and clear constellation map (Figure 2(b) inset). Similar experiments are conducted for 12 Kerr comb lines (three at a time), and a total of 180 CO-OFDM bands are generated and transmitted, adding up to a total capacity of 6.885 Tb/s with a spectral efficiency as high as 2.625 bit/Hz/s (i.e., 3×0.93×16/17×191.25/191.31), compromised mainly by the 7% forward-error correction and 1/16 cyclic prefix overhead. Figure 2(c) shows the estimated SNR and BER for each of the 180 CO-OFDM bands, the majority (175 bands) of which achieves BER below the 7% FEC threshold (Figure2(c)), as well as excellent BER sensitivity at the receiver (Figure2(e)). A few CO-OFDM bands exhibit higher BER due to inferior EO modulation efficient at their wavelengths, which can be readily optimized. These results confirm the capability of our hybrid Kerr-EO frequency comb for coherent terabit data transmission.

As mentioned above, the most pivotal feature of Kerr frequency comb as a multi-tone multi-color laser source is its high coherence and namely, stable mode spacing, which allows the elimination of guard interval between adjacent data channels. For our hybrid Kerr-EO frequency comb, the mode spacing stability of Kerr comb is characterized via the 12.8095 GHz spacing of the edge comb lines from adjacent EO comb families. As shown in Figure 3(a), those edge comb lines, e.g. the 15th and 16th comb lines, achieves comparable transmission performance with the EO comb lines (e.g. comb lines 1 to 14). Indeed, SNR degradations are observed for a few subcarriers near the joint point of two OFDM bands, which is due to the discrepancy between the comb spacing (12.8095 GHz) and the CO-OFDM baud rate (12.75 GHz), as noted above. Nevertheless, this 59.5 MHz mismatch only induces interferences among a couple of edge subcarriers, which do not cause destructive impact on the overall receiving performance of these two CO-OFDM bands [30].

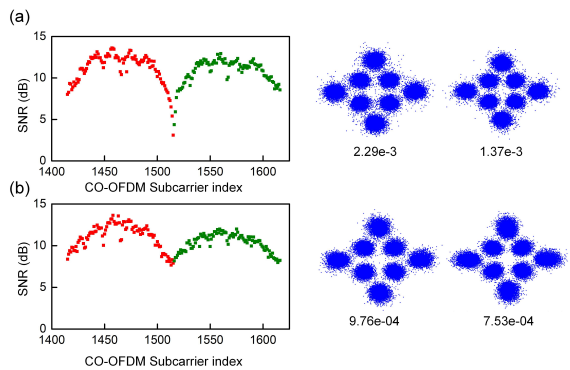

Fig. 3. (a). SNR performance (left panel) and constellation map (right panel) of CO-OFDM bands transmitted using comb line 15 and 16, with 12.8095 GHz mode spacing. The number in right panel gives the BER for each OFDM band. (b). Same as (a) but with 12.7500 GHz comb mode spacing between comb line 15 and 16.

To directly validate the DKS comb stability for seamless CO-OFDM transmission, in a new experiment we deliberately change the 38.25 GHz RF signal to 38.264955 GHz, so as to allow the spacing between the 15$^{th}$ and 16$^{th}$ comb lines (which reflects the stability of Kerr frequency comb) to be exactly 12.75 GHz, and repeat the CO-OFDM transmission test. In such a situation the two CO-OFDM bands are now fully orthogonal, and they consequently demonstrate excellent transmission performance, as shown in Figure 3(b). The mean SNR and BER achieved by the 15$^{th}$(16$^{th}$) comb line is 11.10 dB (10.34 dB) and 9.76×10$^{-4}$ (7.53×10$^{-4}$), respectively. No inter-band interference is captured for all OFDM subcarriers, illustrating that the coherence of DKS comb are sufficient to support CO-OFDM transmission.

Finally, it is worth mentioning that our hybrid frequency comb combines the merits of both Kerr and EO combs, with which we not only overcome the small power problem suffered by low repetition rate DKS combs, but also avoid the necessity to directly generate a broadband EO comb from nonlinear spectrum broadening [31]. Moreover, EO comb provides flexible comb spacing control, which can be a viable feature for superchannel networks. In our experiment, one Kerr comb line is modulated to 15 EO comb lines, such modulation enables us to link the 191.31 GHz cavity free spectral range to the 20 GHz instrument bandwidth of our eAWG. Nevertheless, modulation complexity (i.e. number of EO comb lines) can be reduced by using either faster eAWG or smaller cavity free spectral range as long as each individual Kerr frequency comb line can be effectively amplified while maintaining high OSNR for transmission [32].

**Funding.** The authors acknowledge support from NFSC grant 61705033, 61405024, Office of Naval Research N00014-14-1-0041, the National Science Foundation 17-41707 and 15-20952.